\documentclass[fleqn,usenatbib,useAMS]{mnras}

\usepackage{newtxtext,newtxmath}
\usepackage{natbib} 

\usepackage[utf8]{inputenc} 
\usepackage[T1]{fontenc}

\usepackage{graphicx} 
\usepackage{amsmath}

\usepackage{xspace}	
\usepackage{color}
\usepackage{lineno}
\usepackage{tablefootnote}
\usepackage{flushend}
\usepackage{cuted}
\usepackage{natbib}
\usepackage{threeparttable}
\usepackage{xcolor}
\usepackage{float}
\usepackage{dcolumn}
\usepackage{multicol}
\usepackage{adjustbox}
\usepackage{multirow}
\usepackage{caption}
\usepackage{supertabular}
\usepackage{subcaption}
\usepackage{longtable}
\usepackage{caption}

\newcommand{\caltech}{1}
\newcommand{\iitbmech}{2}
\newcommand{\iitb}{3}
\newcommand{\nasa}{4}
\newcommand{\iia}{5}
\newcommand{\heid}{6}

\newcommand{\sw}[1]{\texttt{#1}}

\newcommand{\arcsecs}{{$^{\prime\prime}$}\xspace}

\graphicspath{{./}{figures/}}

\usepackage{graphicx}

\title[Astreaks]{Astreaks: Astrometry of NEOs with trailed background stars}

\author[K. Sharma et al.]
{Kritti Sharma,$^{\caltech, \iitbmech}$\thanks{E-mail: kritti@caltech.edu}
Harsh Kumar,$^{\iitb}$\thanks{E-mail: harshkumar@iitb.ac.in}
Harsh Choudhary,$^{\iitb}$
Varun Bhalerao,$^{\iitb}$
Vishwajeet Swain,$^{\iitb}$
Bryce Bolin,$^{\caltech, \nasa}$
\newauthor
G.C. Anupama,$^{\iia}$
Sudhanshu Barway,$^{\iia}$
Simran Joharle,$^{\heid}$
Vedant Shenoy$^{\iitb}$
\\
${\caltech}$~Division of Physics, Mathematics, and Astronomy, California Institute of Technology, Pasadena, CA 91125, USA\\
${\iitbmech}$~Department of Mechanical Engineering, Indian Institute of Technology Bombay, Powai, 400 076, India \\
${\iitb}$~Physics Department, Indian Institute of Technology Bombay, Powai, 400 076, India\\
${\nasa}$~Goddard Space Flight Center, 8800 Greenbelt Road, Greenbelt, MD 20771, USA, and NASA Postdoctoral Program Fellow \\
${\iia}$~Indian Institute of Astrophysics, 2nd Block 100 Feet Rd, Koramangala Bangalore, 560 034, India\\
${\heid}$~Heidelberg University, Grabengasse 1, 691 17, Heidelberg, Germany\\
}
\date{Accepted 2023 June 28. Received 2023 June 28; in original form 2023 April 27}

\pubyear{2023}

\begin{document}
% \linenumbers
\label{firstpage}
\pagerange{\pageref{firstpage}--\pageref{lastpage}}
\maketitle

%%%%%%%%%%%%%%%%%%%%%%%%%%%%%%%%%%%%%%%%%%%%%%%%%%%%%%%%%%%%%%%%%%%%

\begin{abstract}
The detection and accurate astrometry of fast-moving near-Earth objects (NEOs) has been a challenge for the follow-up community. Their fast apparent motion results in streaks in sidereal images, thus affecting the telescope's limiting magnitude and astrometric accuracy. A widely adopted technique to mitigate trailing losses is non-sidereal tracking, which transfers the streaking to background reference stars. However, no existing publicly available astrometry software is configured to detect such elongated stars. We present \sw{Astreaks}, a streaking source detection algorithm, to obtain accurate astrometry of NEOs in non-sidereal data. We validate the astrometric accuracy of \sw{Astreaks} on 371 non-sidereally tracked images for 115 NEOs with two instrument set-ups of the GROWTH-India Telescope. The observed NEOs had V-band magnitude in the range [15, 22] with proper motion up to 140\arcsecs/min, thus resulting in stellar streaks as high as 6.5$^\prime$ (582 pixels) in our data. Our method obtained astrometric solutions for all images with 100\% success rate. The standard deviation in Observed-minus-Computed (O-C) residuals is 0.52\arcsecs, with O-C residuals <2\arcsecs(<1\arcsecs) for 98.4\% (84.4\%) of our measurements. These are appreciable, given the pixel scale of $\sim$0.3\arcsecs and $\sim$0.7\arcsecs of our two instrument set-ups. This demonstrates that our modular and fully-automated algorithm helps improve the telescope system's limiting magnitude without compromising astrometric accuracy by enabling non-sidereal tracking on the target. This will help the NEO follow-up community cope with the accelerated discovery rates and improved sensitivity of the next-generation NEO surveys. \sw{Astreaks} has been made available to the community under an open-source license.
\end{abstract}


%%%%%%%%%%%%%%%%%%%%%%%%%%%%%%%%%%%%%%%%%%%%%%%%%%%%%%%%%%%%%%%%%%%%

\section{Introduction} 
\label{sec:intro}

% \outline{
% \begin{itemize}
%     \item \textbf{General introduction to the field:} Why do people study fast-moving objects (including NEOs) and comets? What is the scientific outcome?
    
%     \item \textbf{Motivation:} Why is it important to study NEOs in particular? Then say that there are surveys that discover these NEOs and emphasise the importance of follow-up observations. Build-up how GIT is an awesome tool for a NEO follow-up program.
    
%     \item \textbf{What is our task at hand?} Introduce the problem statement. Summarise the existing approaches. What challenges did we face?
    
%     \item \textbf{Summarise the results:} What is our approach to solving this problem? Why this approach? Main results/highlights.
    
%     \item \textbf{Outline of the paper}
% \end{itemize}
% }

Small Solar System bodies are remnants of the formation stage of the Solar System. These bodies encompass small natural objects like Near-Earth Objects (NEOs), main-belt asteroids, trans-Neptunian objects, and various other smaller groups of asteroids and comets. Their properties, such as size, shape, rotation and surface composition, are the result of collisions and dynamical evolution that has led to their formation. Asteroid science encompasses studies ranging from formation mechanisms, population, collisional evolution, orbital dynamics, compositional properties and physical mechanisms such as Yarkovsky and YORP effects~\citep{2015aste.book.....M}. Out of all small solar system bodies, NEOs are of particular interest to the planetary science community, not only from a scientific perspective but also because of the hazardous consequences of their impacts on civilization~\citep{2013A&ARv..21...65P}. The planetary defence missions to devise efficient mitigation strategies critically depend on timely detection and accurate knowledge of orbits and physical properties of these potentially hazardous asteroids~\citep{2022LPICo2681.2035R, 2022PSJ.....3..148N}. The Double Asteroid Redirection Test (DART) mission is a planetary defence-driven test of technologies for preventing an asteroid's impact on Earth, aimed at demonstrating the kinetic impactor technique for changing the motion of the moonlet of asteroid (65803) Didymos~\citep{2021PSJ.....2..173R, 2020Icar..34813777N, 2023arXiv230302077T,2023arXiv230302248T}. Several survey telescopes scan the night sky daily and report these NEO candidates to Minor Planet Center (MPC). With the current ground-based facilities, NEOs are typically discovered at a distance of less than $1$~AU~\citep{Jedicke2016}. At this distance, their apparent rate of motion can be high, thus challenging their follow-up and recovery.

Most of the telescope facilities operate in the sidereal mode during their regular operations. The high apparent motion of NEOs with respect to the far-situated background astrophysical objects results in a streak in the astronomical images. The quest of discovering and characterizing the NEOs is steered by robotic survey telescopes with wide fields of view~\citep{2021BAAS...53d.241S}. These surveys detect NEOs by their apparent motion between successive exposures and submit the NEO candidates to the MPC. Usually, this discovery data consists of at least two detections, known as a ``tracklet''~\citep{2007Icar..189..151K}. The short observation arcs from the discovery data result in high uncertainties in the preliminary orbit estimate, which could lead to hundreds of arcseconds of uncertainties in the sky positions within a few hours after the discovery for the fastest, nearby, objects. Therefore, well-timed subsequent follow-up observations by meter-class telescopes like GROWTH-India Telescope~\citep[GIT; ][]{2022AJ....164...90K}, with relatively wide fields of view are needed to affirm the candidacy of an NEO.

As discussed, due to sensitivity limitations of the current survey facilities, most of the candidates are discovered at a distance $\lesssim$~1~AU on their discovery apparition~\citep{Jedicke2016}. This results in a high apparent motion $\gtrsim10$\arcsecs/min that degrades the signal-to-noise ratio (SNR) of such candidates as the photons spread over a larger number of pixels in the form of a streak~\citep{2014ApJ...782....1S}. Bright candidates create bright streaks that are detectable in such images, but fainter objects get blended into the background, causing a reduction in the detection limit of these objects compared to other sidereal targets. Furthermore, most of the NEO discovery engines like Zwicky Transient Facility \citep[ZTF;][]{Bellm_2018}, Catalina Sky Survey \citep[CSS;][]{2018DPS....5031010C}, Asteroid Terrestrial-impact Last Alert System \citep[ATLAS;][]{2018PASP..130f4505T} and Panoramic Survey Telescope and Rapid Response System \citep[Pan-STARRS;][]{2016arXiv161205560C} are located in North America and Hawaii island, which means that there are large geographical gaps between the discovery system and follow-up systems, typically in North and South America. Together, these factors make the recovery of NEOs in sidereally-tracked data quite challenging. 

The 70-cm fully-robotic GIT was set up as a part of the international collaboration ``Global Relay of Observatories Watching Transients Happen''~\citep[GROWTH;][]{2019PASP..131c8003K}. GIT (MPC Observatory Code: N51), located at the Indian Astronomical Observatory, Hanle-India, has a 16.8-megapixel sensor which provides a large field of view of 0.5~$\mathrm{deg}^2$, thus making it an excellent tool of the trade for NEO follow-up campaigns. Furthermore, its geographical location at Hanle is on the opposite side of Earth to the major NEO discovery engines, allowing us to observe the candidates before the positional uncertainties blow up to unmanageable scales~\citep{2021EPSC...15..378S}.

The next generation NEO survey programs like NEO Surveillance Mission \citep[NEOSM;][]{2020DPS....5220803G} and Vera C. Rubin Observatory's Legacy Survey of Space and Time \citep[LSST;][]{2021BAAS...53d.236V} are expected to increase the number of NEOs discoveries with absolute magnitude, H<22~mag by 26\% (corresponding discovery rate $\sim2\times$) against the existing surveys over a period of 10 years of its planned operations~\citep{2018Icar..303..181J}, along with the need of deep limiting magnitude follow-up $r \sim 24.5$~mag~\citep{2017arXiv170506209C}. As typical depth probed by small telescopes in a reasonable sidereal exposure is around $V \sim 22$~mag, LSST and NEOSM together will increase the depth requirements of follow-up facilities by couple of magnitudes~\citep{Seaman2021NEO}. This indicates an emerging need to improve the limiting magnitude of the follow-up telescope campaigns of such fast-moving NEOs. The NEO follow-up community can mitigate the damaging effects of trailing losses by tracking at the rate of NEO's apparent motion instead of the stars with existing facilities, thus preserving the point spread function of the NEO~\citep{2012PASP..124.1197V}. This non-sidereal tracking improves the system's limiting magnitude for fast-moving objects by transferring the trailing to background reference stars. However, the astrometric measurement with the trailing stars is a challenging task with existing software that assume largely symmetric point spread functions. These software are not designed to obtain astrometry with elongated reference stars. Moreover, most of these software require human intervention to perform various tasks, which will be a bottleneck in the follow-up campaigns with the advent of the next-generation survey telescopes. Therefore, in order to complement the self-follow-up strategies of these surveys, we look forward to robust astrometry techniques to support the improvement in NEO orbital catalogues.

Here, we present a novel source detection algorithm, \sw{Astreaks} --- Astrometry with Streaking Stars, to obtain an astrometry solution for astronomical images with elongated reference stars\footnote{This work was also presented at the European Planetary Science Congress 2021~\citep{2021EPSC...15..378S}.}. Developed for GIT, \sw{Astreaks} achieves the goal of accurate astrometry using the image segmentation technique, implemented using publicly available astronomy python packages. The pipeline has been validated on two different instrument set-ups and achieves sub-arcsecond astrometric accuracy.
%\footnote{We compute this as the standard deviation of the Observed-minus-Computed residuals of the orbital fit calculated by using our astrometric measurements and existing observations at MPC database as of the time of our observations.}. 
The operations of our fully-automated pipeline can be easily transferred to any other telescope system due to its modular nature with minor modifications in the configuration file. This article is organized as follows. In \S~\ref{sec:background}, we review the loss in astrometric accuracy and limiting magnitudes due to trailing losses, existing methods to recover these trailing losses, and the benefits of non-sidereal tracking. We elaborate on the analysis framework of our elongated source detection algorithm and validate the astrometric accuracy of \sw{Astreaks} in \S~\ref{sec:astreaks_workflow}. Finally, we conclude with a summary and future outlook in \S~\ref{sec:conclusion}.
%%%%%%%%%%%%%%%%%%%%%%%%%%%%%%%%%%%%%%%%%%%%%%%%%%%%%%%%%%%%%%%%%%%%

\section{Sidereal \& Non-Sidereal Observations} 
\label{sec:background}

% \begin{figure} 
% \centering
% \includegraphics[width = \columnwidth]{figures/O-C_Residuals-all_observed_positions_of_Comets-proper_motion_logscale.png}
% \caption{The O-C residuals in right ascension (RA) and declination (DEC) for 331 positions of 44 slow-moving minor planets (including comets and main belt asteroids) were observed using sidereal tracking. The proper motion of these minor planets is in the range of $0.18 - 2.36$\arcsecs/min. The standard deviation of O-C residuals for their astrometric measurements is 0.41\arcsecs. We observe an increase in O-C residuals with an increase in the target's proper motion due to the elongation of target objects in long sidereal exposures. \hk{We should remove this graph as discussed last time. Let me know if the final decision on this.}}
% \label{fig:o-c_sidereal_obs}
% \end{figure}

\begin{figure} 
\centering
    \begin{tabular}{c}
        \includegraphics[width = \columnwidth]{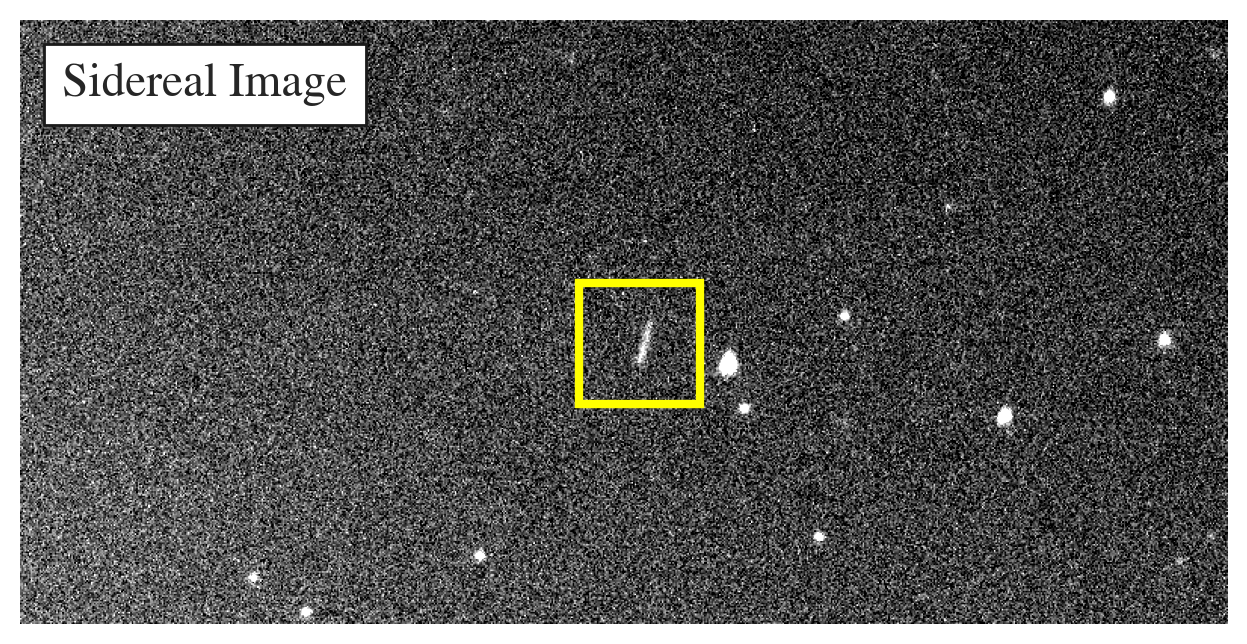} \\
        % \vspace{-5mm}
        \includegraphics[width = \columnwidth]{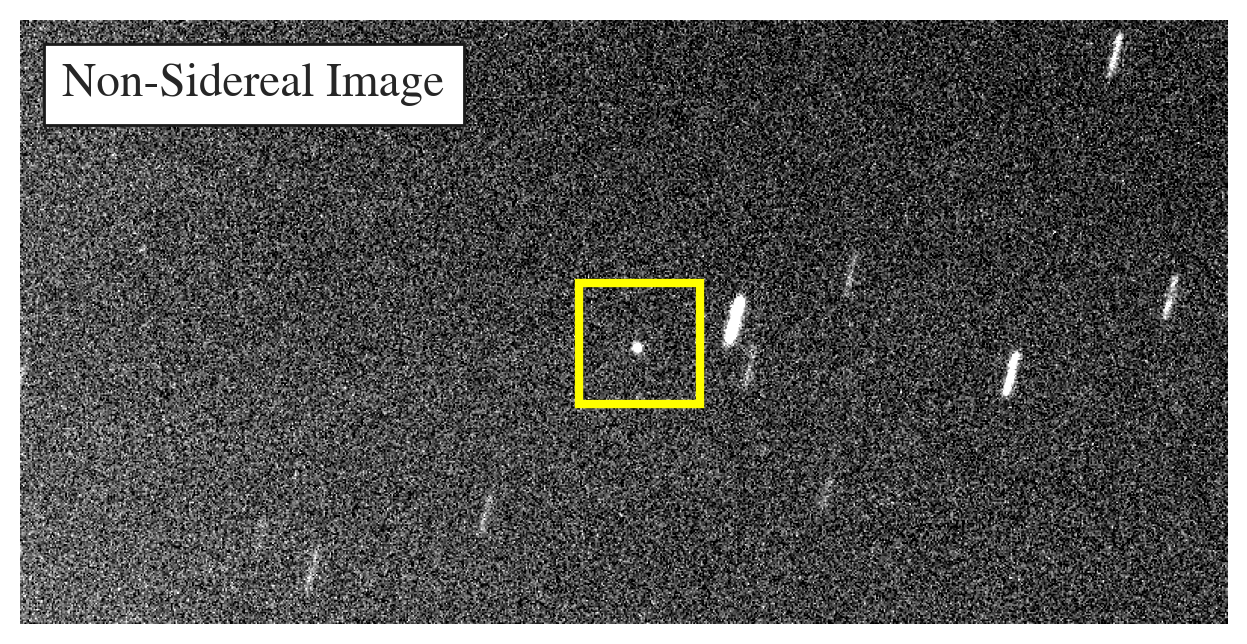}
    \end{tabular}
\caption{Comparison of sidereal (upper panel) and non-sidereal (lower panel) images for 2000 NM, a NEO with an apparent motion of $\sim$4\arcsec/min during our observations. The target streaks in the sidereal exposure and this streaking gets transferred to background reference stars in the non-sidereal exposure. The resulting loss in the signal-to-noise ratio of the target in sidereal exposure is $\sim 20$\%, primarily due to trailing losses.}
\label{fig:sid_vs_non_sid_exp_with_streaking_stars}
\end{figure}

Sidereal observations are where we track the motion of the stars across the sky. To perform the observations in sidereal mode, the telescope motion is corrected for the motions of the stars around the north celestial pole. This helps us keep the star's image focused on the same group of pixels even when the stars are moving around the pole. However, solar system objects usually have a different sky plane motion compared to the stars. Therefore, unlike stars, these objects do not result in the typical 2D Gaussian profile in an image. The minor planets are an example of such objects which leave trails in sidereally-tracked exposures due to their significant apparent rate of motion. The light from these objects spread over many pixels along their motion, causing streaks in the images (Figure \ref{fig:sid_vs_non_sid_exp_with_streaking_stars}). These streaks result in a decrease in the signal-to-noise ratio of the source, which leads to two effects -- a decrease in the detection sensitivity and a decrease in astrometric accuracy for such objects.

% \todo{update para}To demonstrate the damaging effects of sidereal observations for such objects, we calculated the astrometric uncertainty for 18 main belt asteroids and 26 comets using sidereally-tracked data. The astrometric solution was obtained using an offline engine of astrometry.net \citep{Lang_2010}. The Observed-minus-Computed (O-C) residuals in right ascension and declination were computed using the \sw{find\_orb}~\citep{findorb} software by appending all observations of these targets at the MPC database. The O-C residuals are a direct measurement of the astrometric uncertainty and are demonstrated in Figure~\ref{fig:o-c_sidereal_obs} for these 44 minor planets. The standard deviation of these O-C residuals for 331 observed positions of these minor planets is 0.41\arcsecs. As can be observed in the figure, these O-C residuals of the objects increase with the proper motion of the targets due to the elongation of the streak in the images. 

As discussed above, the trailing of minor planets in the sidereal image spreads the flux from the object over a larger area. This causes a reduction in the target's apparent magnitude per unit area and hence signal-to-noise ratio~\citep{2004SoSyR..38..241K}. To demonstrate these effects, we observed the NEO, 2000 NM in sidereal as well as non-sidereal mode, where we tracked the motion of the NEO rather than the stars in the background. We conducted these observations on 2022-08-04 at 16:31:57.0 UTC, where we took 5~minute-long sidereal and non-sidereal exposures. Figure~\ref{fig:sid_vs_non_sid_exp_with_streaking_stars} shows the trailing loss in sidereal exposure for the NEO when compared against its non-sidereal exposure. The target was moving at an apparent rate of 4.01\arcsecs/min, and the typical seeing during our observations was 2.65\arcsecs. Therefore, the target streaked in the sidereal exposure with an aspect ratio of $\ell/w \sim 3.8$. The signal-to-noise ratio of the target's streak in sidereal exposure is 13.0. A similar measurement on the non-sidereal exposure for the target yields a signal-to-noise ratio of 16.4. Therefore, we observe a $\sim 20$\% loss in the signal-to-noise ratio of the target in the sidereal exposure compared to non-sidereal exposure. This loss of signal-to-noise ratio of the trailed object reduces the limiting magnitude of the system for such fast-moving asteroids, thus reducing the probability of detection for these NEOs in sidereally-tracked astronomical images~\citep{1991AJ....101.1518R}.

Along with the decrease in the limiting magnitude of the system, these streaks in the sidereal observations also affect the astrometric measurements. Astrometric measurements in images with significant elongations require non-trivial techniques like ``trail fitting'' \citep{2012PASP..124.1197V}.
% In such cases, it is inappropriate to do measurements for minor planets with significant elongation in sidereal exposures without ``trail fitting'', which is a non-trivial technique~\citep{2012PASP..124.1197V}.
Furthermore, it has been tested that the astrometric accuracy of the trail-fitting algorithm is compromised at low SNR~\citep{2012PASP..124.1197V}. Thus, the reduced SNR of the target in sidereally-tracked astronomical images results in increased uncertainty in astrometric measurements of the streaked NEOs. This loss in astrometric accuracy due to trailing losses is even more important for fast-moving NEOs in the discovery apparition since their recovery is dependent on accurate astrometry.

% \improve{The minor planets leave trails in sidereally-tracked long astronomical exposures due to their significant apparent rate of motion. These trails introduce measurable astrometric errors.} \ks{WHY ARE WE SAYING THIS AGAIN?} \hk{This one was from your draft. I forgot to comment as i have rewritten this para.}

\begin{figure} 
\centering
\includegraphics[width = \columnwidth]{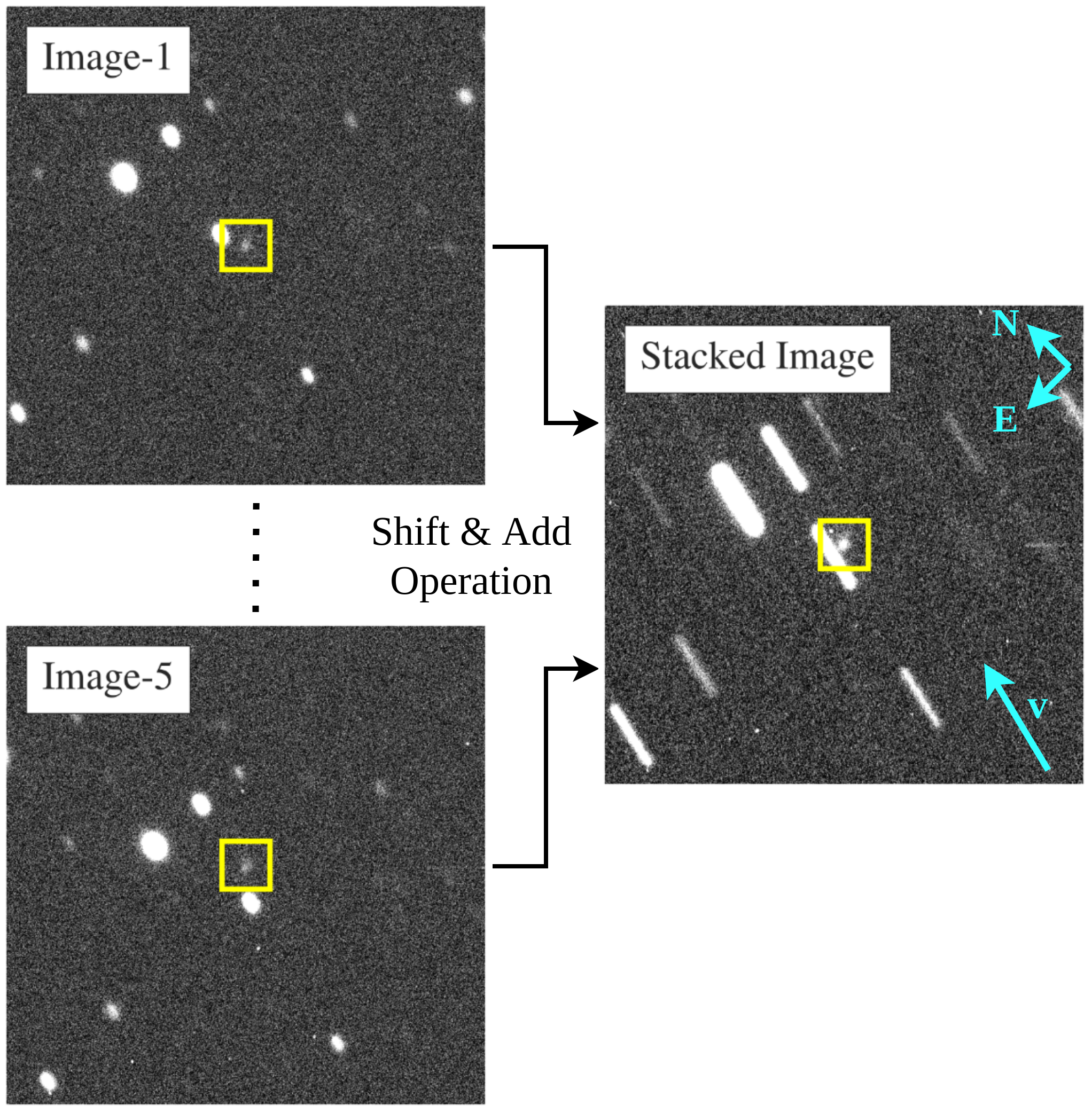}
\caption{Illustration of shift and add operation when implemented on sidereally-tracked data of Comet C/2021 A4 (NEOWISE). The comet had a sky-plane motion of 1.33\arcsecs/min and an apparent magnitude >20 at the time of our observations. We observe that on co-adding the frames with appropriate shifts for the motion of the comet, we get  $\gtrsim$3-fold amplification in the signal-to-noise ratio of the comet in our data as compared to single images.}
\label{fig:shift_and_add_technique}
\end{figure}

\begin{figure*}
\centering
\includegraphics[width = \textwidth]{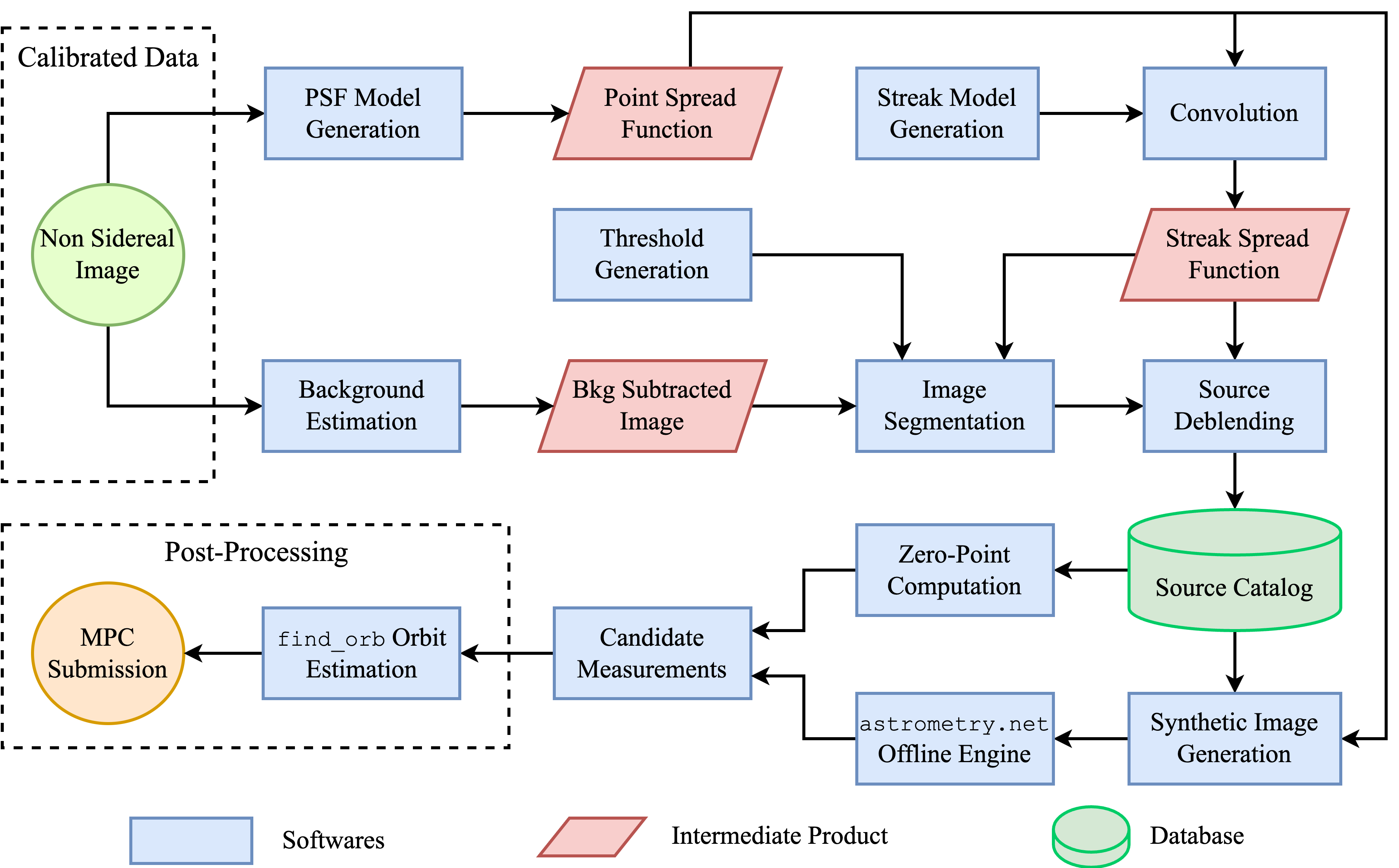}
\caption{Flowchart of the GROWTH-India Astrometry Pipeline, \sw{Astreaks}, for astrometry and photometry of NEOs in non-sidereally tracked data.}
\label{fig:astreaks_workflow}
\end{figure*}

Over the years, several image processing techniques have been proposed to recover the target and improve both astrometry and photometry of the target in sidereally-tracked data. The working principle of a majority of these is based on the ``shift-and-add'' technique to enhance sensitivity ~\citep{1995ApJ...455..342C, 1992AAS...181.0610T, 2010PASP..122..549P}. In this technique, we acquire several sidereally-tracked short exposures of the field, which are then shifted based on the known velocity vector of the minor planet and co-added to improve its signal-to-noise ratio. We illustrate a basic implementation of this technique in Figure~\ref{fig:shift_and_add_technique} on sidereally-tracked data for Comet C/2021 A4~\citep[NEOWISE;][]{MPEC2021-A207}. These observations were conducted at the discovery apparition of this comet and were consequently submitted to MPC. The comet had a proper motion of 1.33\arcsecs/min at a position angle of 342.9$^\circ$ during our observations. The average signal-to-noise ratio in five individual sidereally-tracked images is $\sim$25, which got amplified to $\sim$88 in the stacked image. This improved signal-to-noise ratio enables us to recover faint targets, overcome trailing loss, and confidently detect cometary activity. However, this technique is sub-optimal for fast-moving objects due to its requirement of astrometry solutions for individual images. Such objects require short exposure times to retain their Gaussian point spread function, thus resulting in less reference stars in the field to obtain astrometry solutions. Furthermore, short exposure time causes a significant loss in observing efficiency due to multiple readout cycles. Moreover, this technique requires the exposure to be long enough to detect the faint objects in individual images. This results in measurable elongation in the point spread function of the NEO, thus compromising the astrometric accuracy.

\citet{2012PASP..124.1197V} used an analytical form of trailing function for trail fitting to yield accurate astrometry and photometry. However, this technique has limited usage for astrometry of faint, fast-moving NEOs observed with small telescopes. Moreover, the loss in photometric and astrometric accuracy increases with higher trail aspect ratios and lower signal-to-noise ratios. \citet{2005AJ....130.1951G} developed a matched-filter-based trail detection technique, which attempts to integrate multiple frames by shifting and stacking based on a hypothesis of the target's velocity and then matched-filtering using the hypothesized template. \citet{2008LPICo1405.8388S} proposed a similar velocity matched-filter-based approach to integrate multiple frames, which increases the aggregate signal-to-noise ratio using a multi-hypothesis velocity vector. These methods are computationally expensive due to the huge size of the velocity hypothesis set. Also, the storage of multiple intermediate image products requires a significant amount of computer memory. These techniques have also exhibited high false alarms near the limiting magnitudes of images suggesting that they are not suitable for smaller telescopes as they have low limiting magnitudes~\citep{2008LPICo1405.8388S}. 

\citet{2014ApJ...782....1S} demonstrated an enhanced implementation of the shift-and-add technique called ``synthetic tracking'' to enhance the signal-to-noise ratio of the target and avoid astrometric errors due to trailing losses. This method yields the best results on a large telescope with a CMOS camera for rapid frame acquisition and low readout noise. However, it fails to impress on the CCD camera, where the frame rate is not high enough. Further, this method has limited yields on a small telescope with shallow limiting magnitude. In addition, it requires GPUs for data processing since integrating images with a grid of tracking velocities is computationally expensive. As with other techniques, a drop in the number of reference stars in short exposures compromises the astrometric solution of images. 

\begin{figure*}
\includegraphics[width = \textwidth]{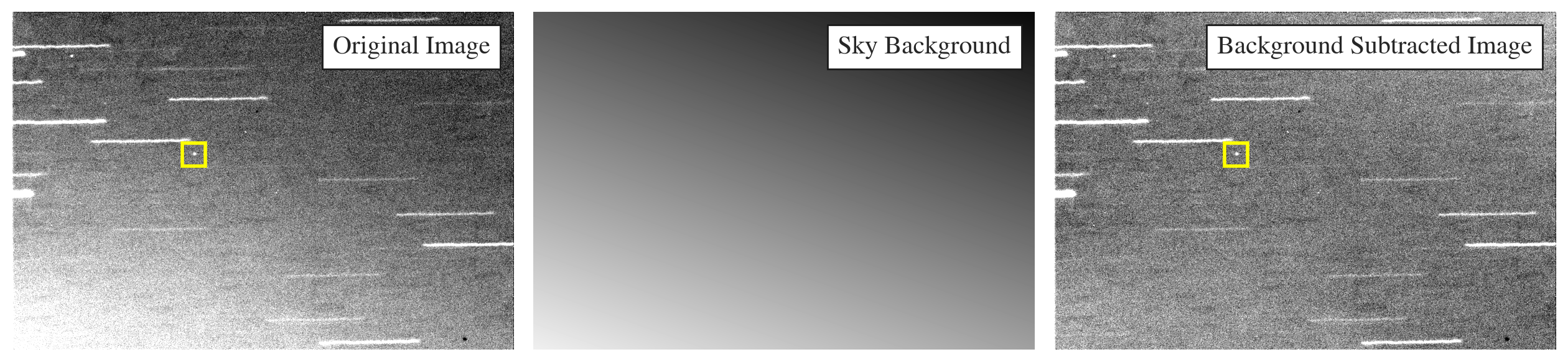}
\caption{Illustration of the sky background estimation technique used in \sw{Astreaks} on non-sidereally tracked image for 2020 XB acquired on 2020-12-01 at 17:09:48 UTC (left panel). A gradient is present in the image because the moon was only 38$^\circ$ from the target} at the time of observations, which is appropriately captured by our sky background estimate (middle panel). Before proceeding, we subtract this sky background from our science image, as displayed in the right panel.
\label{fig:bkg_estimation}
\end{figure*}

\citet{2021AJ....162..250Z} uses an ``image fusion'' technique to obtain images superimposed on background stars and the NEO, followed by fusion of these two superimposed images to get a single image with all sources where the point spread function (PSF) is the same as that of the telescope system (no streaks).
Similar to~\citet{2014ApJ...782....1S}, this method also requires a camera that enables a rapid frame rate. Also, it involves exposure time tweaking based on the apparent limiting magnitude for a particular exposure to identify the minimum exposure time such that the NEO is observable in each frame. The method performs poorly for fast-moving, faint NEO images obtained with small telescopes. 

As discussed above, sidereal data reduction techniques may need high computational power for processing multiple short exposures, as required by techniques such as synthetic tracking, and has limitations depending on the type of method used for processing. To rescue from the trailing losses of the faint, fast-moving NEOs, a widely adopted technique is to use a non-sidereal tracking mode where we track at the apparent rate of motion of the target~\citep{Kamiski2017HighER, 2022arXiv220703914R, 2018SPIE10702E..4RK, 2018SPIE10704E..2HW}. This preserves the PSF of the target NEO. However, since the sky-plane motion of NEOs can be tens to hundreds of arcseconds per minute, long exposures with non-sidereal tracking yield streaking reference stars (for an example, see the bottom panel of Figure~\ref{fig:sid_vs_non_sid_exp_with_streaking_stars}). The conventional astrometry methods based on detecting symmetric, Gaussian-like sources, are inefficient for obtaining astrometry solutions of images with such elongated reference stars. Hence, calculating accurate asteroid positions and magnitudes from non-sidereally tracked images is challenging. In this work, we have developed a new automated pipeline, \sw{Astreaks}, to process such images and obtain accurate astrometry of fast-moving NEOs in non-sidereal images. 
% The pipeline has been developed based on the data obtained by GIT. 
% However, it can be easily tweaked to use on data acquired with any other telescope. The pipeline details are discussed in following section. 

\section{Astreaks Workflow}
\label{sec:astreaks_workflow}

In order to accurately measure the coordinates of the asteroid in non-sidereal images, we need to robustly calculate the astrometric solution for the image based on stellar streaks. We calibrate the images by applying bias correction and flat-fielding, followed by cosmic rays removal, as implemented in default GIT data reduction software~\citep{2022AJ....164...90K}. \sw{Astreaks} achieves this goal by detecting sources using a ``streak spread function'', coupled with an accurate background estimation technique. It then creates a synthetic image which is eventually used to obtain the astrometric solution. The following sections highlight the detailed working of the pipeline.

% As discussed in the previous section, non-sidereal observations for targets with the proper motion of tens of arcseconds per minute lead to streaking background reference stars. The existing source extraction software is not designed to detect sources with hundreds of arcseconds of elongation. The robustness of the source extraction process to generate a catalogue of objects is of utmost importance to get accurate astrometry. This section discusses how the GROWTH-India astrometry pipeline, \sw{Astreaks}, achieves this goal. The pipeline uses a streak spread function-based source detection method with an accurate background estimation technique to generate a synthetic image which is eventually used to obtain the astrometric solution of the image. The following sections highlight the detailed working of the pipeline.

\subsection{Sky Background Estimation}\label{subsec:bkg_estimation}

Accurate background estimation is crucial for detecting faint sources and determining the correct fluxes from each source. The presence of streaking background reference stars affects the background statistics, due to which, the traditional methods fail to get an accurate background map of the image. To estimate the background, we overlay a grid on the entire image and calculate the mode of counts in each of the cells in this grid: giving a nominal background level for each grid point. We note that mode is a better estimator than the median, as there can be a large number of extended sources which add a heavy tail to the histogram of counts in each grid cell. The grid spacing has to be chosen by keeping in mind the presence of extended sources and occasional crowded fields. If cells are too small, extended sources may dominate some cells and the background estimate will be compromised. If the cells are too large, we will be insensitive to background variations at smaller scales. We recommend that the mesh size should be at least greater than the streaks in the image to avoid detecting the streak itself in the background when using mode as an estimator. Future versions of Astreaks will incorporate an automated mesh size based on the tracking rate and exposure time that determines the length of the stellar streaks. Next, we create a smooth background estimate from these background measurements. In our data, we found that using a linear background variation over the entire image gives satisfactory results. Hence, we fit a plane to this background data by least-squares minimization and subtract this from the original (un-gridded) image. Typically the NEOs observed with GIT have a proper motion of the order of $10$\arcsecs/min, and the typical exposure time is 3 minutes. An average pixel scale of $0.5$\arcsecs/pix amounts to streaks of length $\sim 50$ pixels. Testing 371 images of 115 NEOs, we observed that meshes of size $60-100$ pixels work well. 

Figure~\ref{fig:bkg_estimation} shows our sky background estimation technique on the observations of the minor planet 2020 XB. A background gradient is present in the image because the moon was only 38$^\circ$ from the target. This background gradient is appropriately captured in our corresponding sky background estimate. We subtract this sky background from our image before further analysis.

%%%%%%%%%%%%%%%%%%%%%%%%%%%%%%%%%%%%%%%%%%%%%%%%%%%%%%%%%%%%%%%%%%%%

\subsection{Streak Spread Function Model Generation}\label{subsec:SSF}

Streaked sources in the image can be thought of as a convolution of the PSF of the telescope system and the motion vector of the source in the image~\citep{2020ApJ...888...20L}. Therefore, estimating this ``streak spread function'' (SSF) model is necessary to detect the streaking reference stars. 
% \todo{Explain exact steps to VB} 
We measure the full width at half maximum (FWHM) of the streaks perpendicular to the direction of motion of the target, and create a 2-D Gaussian PSF with the same FWHM. The angle of the streak with respect to image orientation is estimated using \sw{opencv} package~\citep{opencvlibrary}. We use \sw{opencv}'s sigma threshold module to detect streaked objects in the image, and from that, we can get the orientation/angle of the streaks. Further, we compute the expected length of the streak using the tracking rate and exposure time of the image. In the process, we generate a horizontal aggregation of point sources spaced by 1 pixel and spanning across the length of the streak. For more smoother sampling of SSF, the aggregation of point sources spacing can be decreased to a sub-pixel level depending upon the available computational power. This line is then convolved with the Gaussian PSF model and rotated by the computed streak position angle to generate the final SSF model
%\hkj{(as obtained previously)} 
of the streaking stars~\citep{2012PASP..124.1197V} using the \sw{Astropy}~\citep{2013A&A...558A..33A, 2018AJ....156..123A} python package.

% \begin{figure}
%     \centering
%     \includegraphics[width =\columnwidth]{figures/fwhm_measurement.png}
%     \caption{}
%     \label{fig:fwhm_from_ns}
% \end{figure}

\begin{figure}
\centering
    \begin{tabular}{cc}
        \includegraphics[width = 0.45\columnwidth]{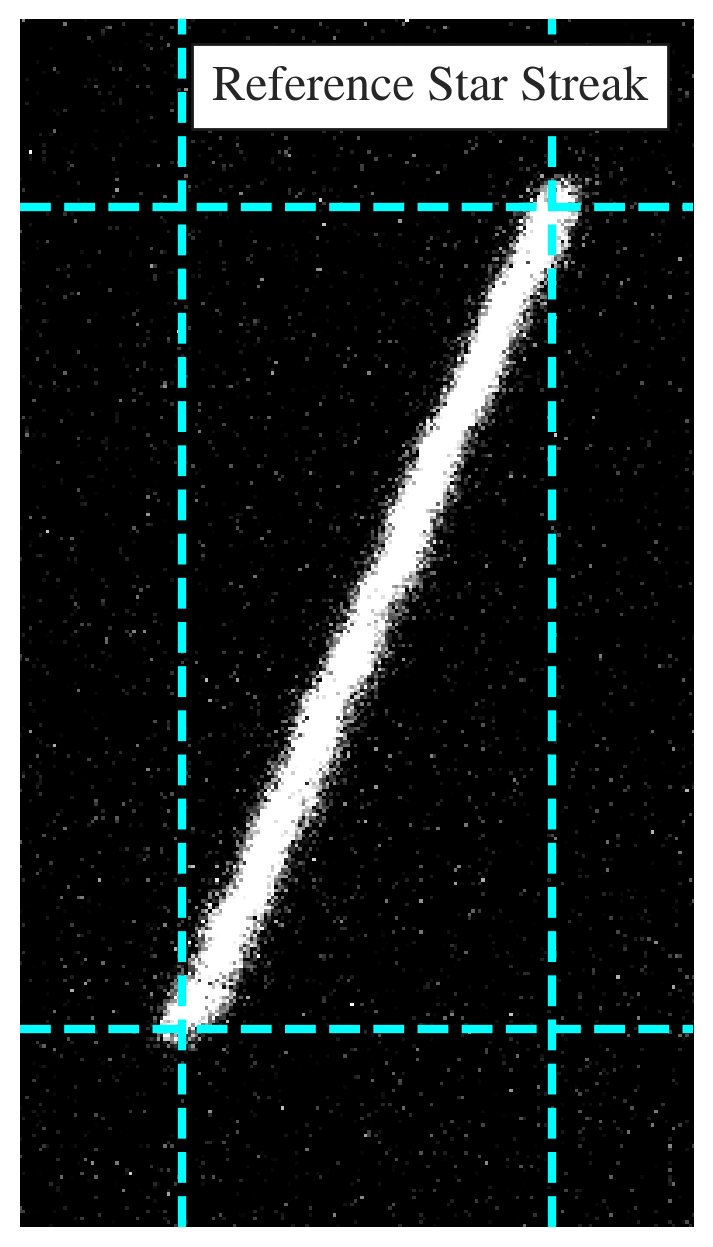} &
        \includegraphics[width = 0.45\columnwidth]{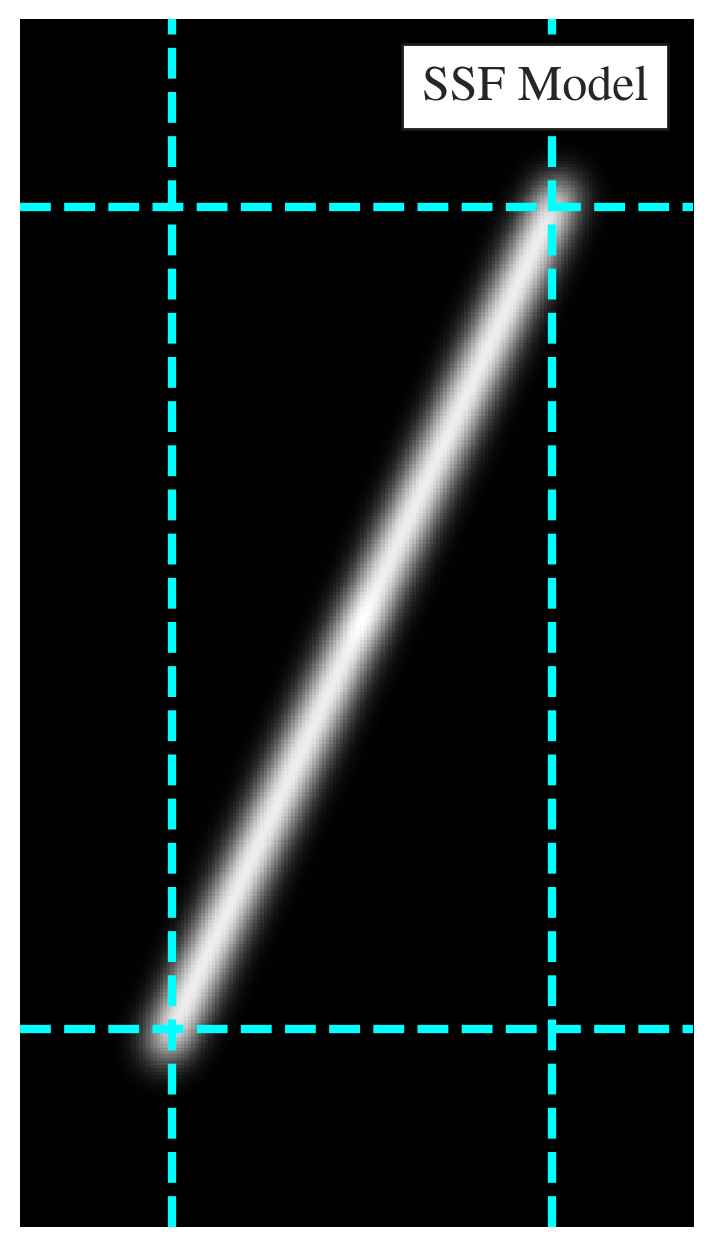}
    \end{tabular}
    \caption{ Comparison of a reference star streak in the 2020 UA1 non-sidereally tracked image (left panel) and the corresponding SSF generated by our pipeline (right panel). The computed start and end points of the streak are marked. The SSF model generated by convolving the PSF model from the non-sidereal image and ideal streak model, which is a line joining these endpoints, appropriately models the reference star streak.}
    \label{fig:SSF}
\end{figure}

Figure~\ref{fig:SSF} compares an original stellar streak in the data acquired for the minor planet 2020 UA1, observed during its discovery apparition, with the corresponding SSF model generated by \sw{Astreaks}. The apparent rate of motion of the target was 39.25\arcsecs/min at the time of our observations, thus leading to 78.5\arcsecs long stellar streaks in the data in exposures of two minutes. The calculated position angle of the target using the velocity position angle (118.2$^\circ$) and telescope position angle ($-52.3^\circ$) is $\sim66^\circ$, which is consistent with our orientation measurement with the sigma-thresholding algorithm. The SSF generated using these parameters accurately represents the stellar streak shown in the figure. 

%%%%%%%%%%%%%%%%%%%%%%%%%%%%%%%%%%%%%%%%%%%%%%%%%%%%%%%%%%%%%%%%%%%%

\subsection{Source Detection}\label{subsec:thresholding_img_seg_deblend}

We use the background subtracted image to detect sources. For this purpose, we generate a $1\sigma$ threshold (a configurable hyper-parameter, see Table~\ref{table:parameters_hyperparameters}) for the image and apply that to detect the sources. 
% \todo{check if this is the ``threshold'' hyperparameter.} \ks{Yes, can be changed in the config file.} 
Such a low threshold has been chosen to detect the faint sources in the image. After that, image segmentation is performed on the background-subtracted image with a normalised SSF model as the kernel using the \sw{photutils} \citep{photutils} python packages. The kernel identifies the pixels with counts greater than the corresponding pixel-wise threshold level. Next, the pipeline identifies the group of pixels corresponding to the same source using the streak length. The segmented sources are then deblended to separate the overlapping sources. The flux inside the elliptical apertures measures the flux from each source. The deblending routine further computes the deblended flux and corresponding flux errors for overlapping sources. The \sw{photutils} python package provides a good measure of flux for deblended sources, hence we chose to use the flux of these sources using the elliptical aperture. We create a ``detected source catalogue'' comprised of the centroids, total flux and flux errors of each source. 
%We implement this routine using the, which uses a combination of multi-thresholding and watershed segmentation to render these jobs.

\begin{figure} 
\centering
\includegraphics[width = \columnwidth]{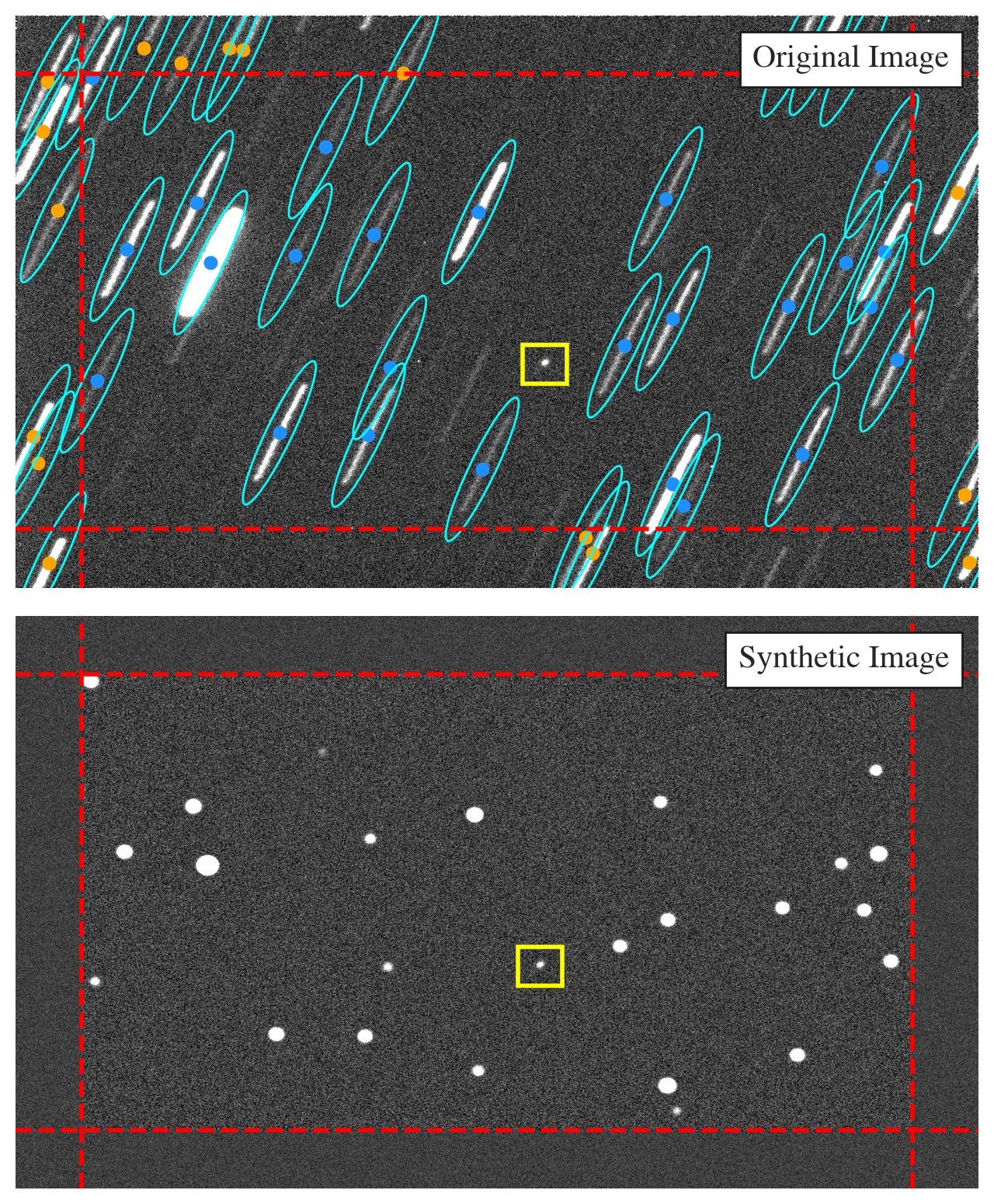}
\caption{Detection of background reference stars using \sw{Astreaks} in a non-sidereally tracked image of 2020 UA1. The detected sources at mid-time of the exposure are marked as dots on top of the corresponding source's streak in the top panel. We replace the streaked sources with their flux-scaled gaussian equivalent sources in a synthetic image, as displayed in the bottom panel and obtain the WCS solution of the resultant synthetic image. During this process we neglect the sources lying close to the image edge (sources marked with orange color in top panel) and perform astrometry of the target at the mid-time of the exposure.}
\label{fig:detected_sources_and_synthetic_img}
\end{figure}

The sources detected by the pipeline in a representative non-sidereally tracked image of 2020 UA1 are highlighted in the top panel of Figure~\ref{fig:detected_sources_and_synthetic_img}. The target NEO is marked with a yellow rectangular box. The blue dots represent the positions of background stars at the mid-time of the exposure that is used as reference stars in a later stage. A few of the streaks lie close to the edge; hence, their centroid location and flux estimation is incorrect as shown by orange marks in the top panel of Figure~\ref{fig:detected_sources_and_synthetic_img}. To remove any biases due to these sources, we remove the streaks whose centroid falls within half the streak length from the edges, denoted by red dashed lines. 

\subsection{Obtaining an Astrometric Solution}\label{subsec:synthetic_img_astrometry_soln}

Next, we use our detected source catalog to generate a synthetic image. A specific advantage of this method is that the synthetic image is broadly similar to the images that would be acquired by the telescope with sidereal tracking. Hence, the usual astrometry pipelines used for that telescope can be used directly for processing this synthetic image. We use a 2-D gaussian PSF, with the same sigma as the ``SSF'' in \S\ref{subsec:SSF} as our model for point sources in the image.
Starting with a blank image, we use this PSF model to inject a source with appropriate flux at the location of each detected source. We do not consider the sources near edges whose properties are known to be inaccurate. The median sky background and its root mean square value estimated from the original image is added to the synthetic image.
%\todo{why does 7b have background noise? how was it created?}.
The resulting synthetic image is shown in lower panel of Figure~\ref{fig:detected_sources_and_synthetic_img}. This synthetic image is then solved for the World Coordinate System (WCS) using the offline engine of \sw{astrometry.net}~\citep{Lang_2010}. We use the WCS of this synthetic image as the astrometric solution for the original image at the mid-time of the exposure.

\subsection{Astrometry, Photometry, and Orbit Determination}\label{subsec:astrometry_find_orb}

The astrometric positions of the target NEO at the mid-time of the exposure are obtained using the WCS of the synthetic image. Target asteroids are often faint even in non-sidereal tracked exposures. Automated searches for it often give false positives. Since the user is aware of the approximate location of the NEO, at this stage, the NEO is first visually identified. Then, its exact position is determined by fitting the 2-D Gaussian PSF from \S\ref{subsec:synthetic_img_astrometry_soln}. We compute the instrumental magnitude of the target using aperture photometry~\citep{photutils}, using the source fluxes measured in the source detection step (\S\ref{subsec:thresholding_img_seg_deblend}). The magnitudes of streaked stars are cross-matched with the Pan-STARRS catalogue ~\citep{2018AAS...23143601F} using \sw{VizieR} query to calculate the zero points. This zero point is directly used to calculate the magnitude of the point source NEO in the image. Our photometry is accurate to $\sim 0.2$ mag and further improvement in the photometric accuracy in progress. We append our astrometric and photometric measurements to the existing observations of the target at MPC and attempt an orbit fit using \sw{find\_orb}~\citep{findorb} to compute the Observed-minus-Computed (O-C) residuals. These O-C residuals are a direct measurement of the astrometric uncertainty.

\subsection{Hyperparameters and Parameters}\label{subsec:hyperparams_params_input_files}

\begin{table*}
    \centering
    \begin{tabular}{|l|l|l|}
        \hline
        \textbf{Parameter Type} & \textbf{Name} & \textbf{Desciption} \\
         \hline \hline
         Hyperparameters & mesh size & The size of each block in the grid used for background estimation. Units: pixels \\
          & pixel scale & The plate scale to convert the size of mesh in data to image pixels.  Units: \arcsecs/pixel\\
          & thresholding level & The assumed background level to threshold the image and identify streaks.\\
          & PSF model size &  The maximum size of the PSF model to be used for generating the SSF. Units: pixels \\
          & gain & The detector gain to scale instrumental counts to the number of electrons per pixel. Units: e$^{-}$/ADU  \\
         \hline 
         Parameters & NEO tracking rate & The sky-plane velocity of the NEO to estimate the streak length. Units: \arcsecs/min \\
         & exposure time & The total non-sidereal exposure time  to estimate the streak length. Units: sec \\
          & velocity position angle & The velocity position angle of the NEO to determine the orientation of the SSF. Units: degrees \\
          & image coordinates of the target & Pixel coordinates of the target in the image to fit a PSF and extract coordinates. \\
         \hline
    \end{tabular}
    \caption{The description of the set of parameters and hyperparameters used by \sw{Astreaks}.}
    \label{table:parameters_hyperparameters}
\end{table*}

\sw{Astreaks} requires various inputs for its modular functioning, which are summarized in Table~\ref{table:parameters_hyperparameters}. 
%\todo{explain all parameters} \ks{(The idea was to explain them in the respective steps where they're used and summarize them in the following text and table.)} 
The astrometry configuration of the pipeline comprises several \textit{hyperparameters} that depend on the instrument properties and are the same for all targets being analysed. Hyperparameters like mesh size, pixel scale, threshold level, the maximum size of the PSF model etc., are required at different stages of source extraction and measurements. An appropriate mesh size depends on the typical length of the streaks in data and plays a primary role in estimating a reliable sky background. The SSF model generation requires pixel scale information to compute streak parameters. Based on the instrument's sensitivity, the threshold and contrast levels are required for source detection. The synthetic image generation requires the maximum PSF model's size which is fixed based on the typical full width at half maximum (FWHM) of the PSF for the telescope system. The photometry of the target requires the camera properties like the gain of CCD, which is typically fixed for a single instrument. All these hyperparameters must be set up for the best results before using the pipeline on a new instrument. 

Apart from the hyperparameters discussed above, \sw{Astreaks} requires a set of \textit{parameters} that must be updated for each target. These include tracking rate, velocity position angle and exposure time which are different based on the target of interest. A combination of these parameters determines the streak parameters and size. The pipeline extracts these parameters automatically using the information stored in the image headers and performs the astrometric operations on the image. Further, the pipeline can generate an automated MPC report once we specify the pixel coordinates of the target.

\subsection{Astreaks Validation}\label{subsec:astreaks_val}
\sw{Astreaks} has been developed on data acquired with GIT. For this purpose, we used observations of 115 NEOs covering a wide range of proper motion. To test the robustness of \sw{Astreaks}, we acquired data on two CCD camera set-ups: Apogee KAF3200EB (Apogee hereafter) and Andor iKon-XL 230 4k back-illuminated CCD (Andor hereafter). As both cameras have different properties, they provide a nice framework to test the reliability and modularity of \sw{Astreaks}. Apogee camera has a narrow $\sim 11^\prime \times 7.5^\prime$ field of view (FOV) camera with a pixel scale of 0.307\arcsecs/pix. On the other hand, Andor has a $\sim 45.87^\prime \times 45.74^\prime$ FOV with 0.67\arcsecs pixel scale.

% \sw{Astreaks} has been developed and validated on GIT observations of 115 NEOs with a huge range of proper motion. GIT began observations of small solar system bodies in September 2020 in coordination with GROWTH using Apogee KAF3200EB (Apogee hereafter) and Andor iKon-XL 230 4k back-illuminated CCD (Andor hereafter) cameras. Both cameras have quite different properties where Apogee is narrow $\sim 11^\prime \times 7.5^\prime$ FOV camera sampled at 0.307\arcsecs/pix resolution, while Andor has a 0.7\degr field of view at 0.67\arcsecs resolution. This provides a nice framework to validate the performance of the pipeline on two different set-ups.

% \todo{update numbers} \hk{Kritti Confirmed all numbers are up to date} \ks{(Note: Does not include observations after November'2022, if there were any.)}

The data used for \sw{Astreaks} development and validation was acquired from September 2020 to December 2022. 64\% of these data are wide field images acquired with Andor camera, the rest were  obtained with the narrow FOV Apogee Camera. A total of 18 NEOs were observed with the Apogee camera, amassing 133 non-sidereal images during the initial phase of \sw{Astreaks} development.
% An observation of 18 NEOs were acquired on Apogee camera containing 133 non-sidereally images in order to develope the \sw{Astreaks}. 
These targets had a proper motion in the range of 4 - 120\arcsecs/min at the time of observation, resulting in 14 - 140\arcsecs long streaks of reference stars in exposures up to 5~minutes. Using \sw{Astreaks}, we obtain astrometric measurements of these NEOs following the procedure outlined in \S\ref{sec:astreaks_workflow}. The O-C residuals in right ascension and declination were computed using the \sw{find\_orb}~\citep{findorb} software by appending all observations of these targets at the MPC database. These O-C residuals have a standard deviation of 0.41\arcsecs, which qualifies the $<2$\arcsecs quality of measurements criterion, as demanded by MPC~\footnote{MPC's Guide to Minor Body Astrometry: \url{https://cgi.minorplanetcenter.net/iau/info/Astrometry.html}}.

\begin{figure}
    \centering
    \includegraphics[width =\columnwidth]{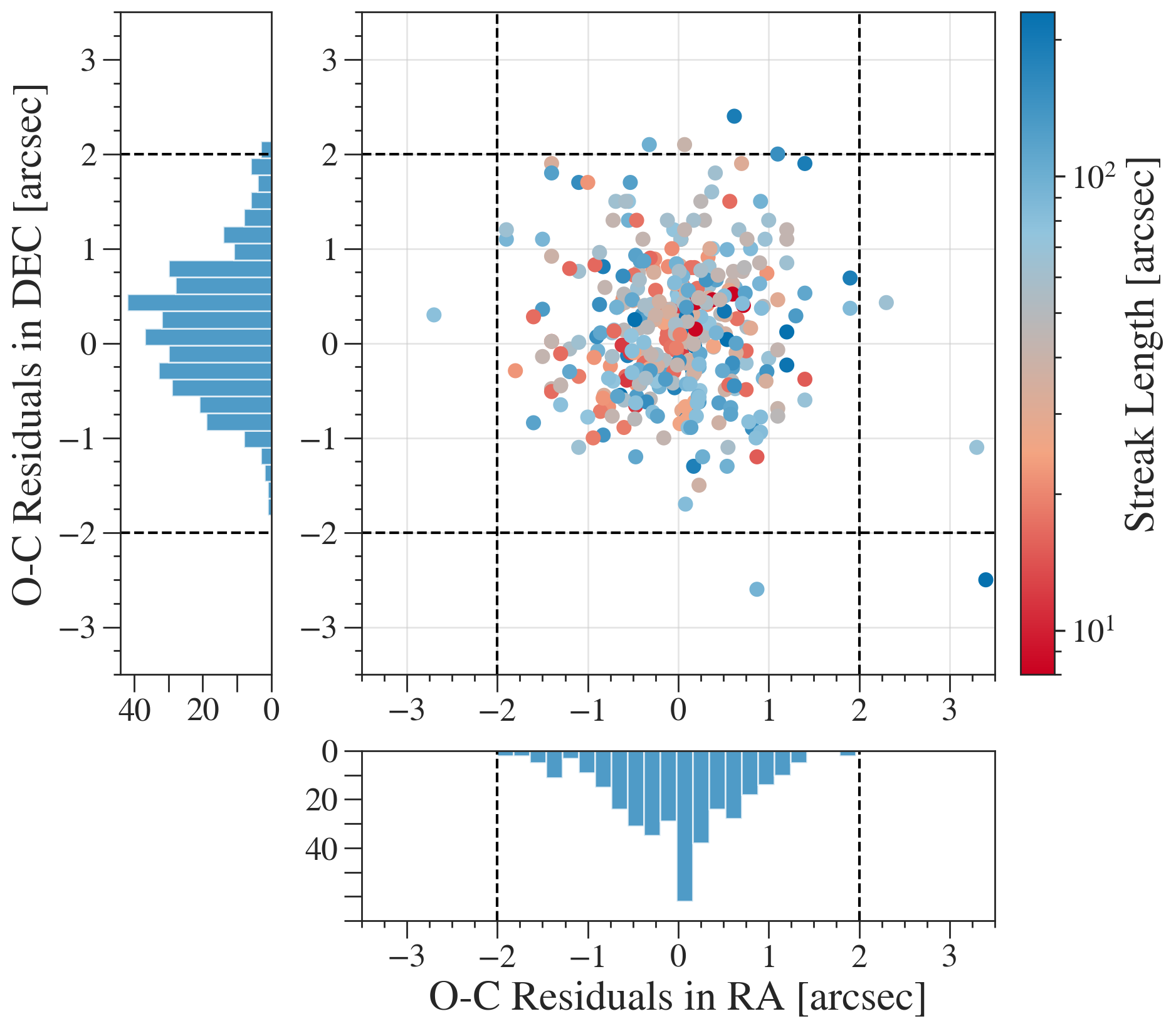}
    \includegraphics[width =\columnwidth]{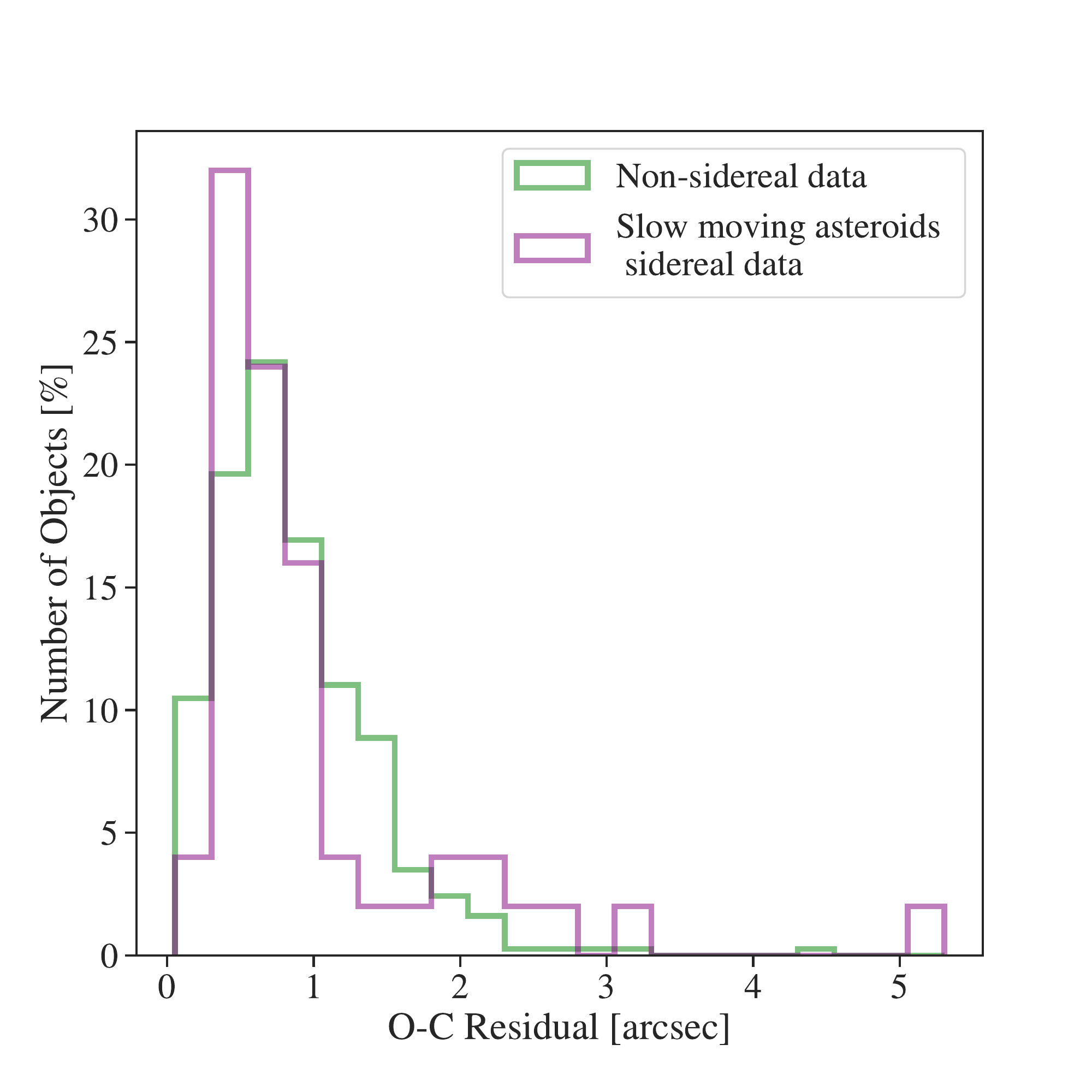}
    \caption{Validation of the astrometric accuracy of \sw{Astreaks}. Upper panel: Validation using O-C residuals in right ascension (RA) and declination (DEC) for 371 positions of 115 NEOs observed with GIT using non-sidereal tracking. The standard deviation of all these measurements is 0.52\arcsecs.
    Lower panel: Comparison of O-C residuals from \sw{Astreaks} reduction of non-sidereal data and slow-moving asteroids in sidereal images reduced using standard procedures. The performance of Astreaks on non-sidereal data is similar to that of normal astrometric measurements from sidereal images.}
    \label{fig:o-c_non_sidereal_obs}
\end{figure}

Once the working and scientific reliability were established with Apogee data, we used 238 non-sidereally tracked images for 96 NEOs acquired with Andor camera data to test the performance of the pipeline. By virtue of the modular nature of our pipeline, we could easily shift from the Apogee to Andor instruments by a simple tweak of  hyperparameters, (see \S~\ref{subsec:hyperparams_params_input_files}). The NEOs had a proper motion in the range of 1- 45\arcsecs/min, resulting in 8--396\arcsecs long stellar streaks in exposures of 1--9~minutes. The computed O-C residuals for astrometric measurements using \sw{Astreaks} have an standard deviation of 0.58\arcsecs. The O-C residuals in RA and declination of all observed positions of 115 NEOs (371 images) are shown in the upper panel of Figure~\ref{fig:o-c_non_sidereal_obs}. More than 84.4\% of our measurements have O-C residuals $<1$\arcsecs, with 98.4\% measurements exhibiting a $<2$\arcsecs astrometric accuracy. We found that the astrometric O-C residuals do not correlate with the streak length (Pearson r value: 0.0074).
%with streak lengths as large as $\sim 6^\prime$, amassing

% \begin{figure}
%     \centering
%     \includegraphics[width =\columnwidth]{astreaks_vs_astrometrica.png}
%     \caption{Comparison of sources detected by \sw{Astreaks} and \sw{Astrometrica}. The markers represent the location of the streaks detected by the two software. We observe that \sw{Astreaks} is capable of detecting fainter sources, compared to \sw{Astrometrica}, which even misses bright sources in the field despite manual cross-matching.}
%     \label{fig:astreaks_vs_astrometrica}
% \end{figure}

We acknowledge that \sw{Astrometrica} software is not designed to detect sources and obtain astrometry solutions with significant elongation in reference stars~\citep{astrometrica}. However, since no streak-analysis packages are publicly available, we compare our source detection capability and astrometric accuracy with \sw{Astrometrica}. We illustrate this comparison on a representative example of non-sidereally acquired data for 2020 UA1. We obtained a total of 9 exposures for 2020 UA1 among which only three images were astrometrically solved with manual cross-match using \sw{Astrometrica}. On the other hand, \sw{Astreaks} successfully solved all nine images with the algorithm discussed in \S\ref{sec:astreaks_workflow}. Furthermore, the number of sources detected by the \sw{Astrometrica} software is less than that of \sw{Astreaks}. Despite a manual cross-match, many bright sources are missed by \sw{Astrometrica}, potentially responsible for the failure of the astrometric solution on the other six images. The number of sources plays a direct role in the quality of the astrometry solution and is higher for \sw{Astreaks} due to the detection of a large number of sources as compared to \sw{Astrometrica}. The standard deviation of astrometric measurements on nine images using \sw{Astreaks} (0.2\arcsecs) is better than the astrometric accuracy on three images using \sw{Astrometrica} (0.3\arcsecs), thus validating that our algorithm is more accurate and robust for obtaining astrometric solutions of non-sidereal images.

As a final test, we also compare the astrometric capabilities of \sw{Astreaks} with standard astrometry procedures on sidereal data. The total astrometric uncertainty for NEOs may include a component from the orbital uncertainty. Hence, for a fair comparison, we use a sample of 50 sidereally acquired images of six slow-moving asteroids with apparent motion less than 0.58~arcsec/min during the exposures. These targets appeared as a point source in our astronomical images of exposure times less than 3 min, given the typical FWHM of PSF in sidereal data of $\sim 3$\arcsec. Astrometry on these images was performed using our usual, standard astrometry procedures for sidereal data, as described in \citet{2022AJ....164...90K}. A comparison of O-C residuals in sidereal data reduced following standard astrometry procedures and non-sidereal data reduced using \sw{Astreaks} is shown in the lower panel of figure~\ref{fig:o-c_non_sidereal_obs}. We observe that the performance of \sw{Astreaks} on non-sidereal data is comparable to normal astrometry methods on sidereal data.

\section{Conclusion and future prospects for \sw{Astreaks}} \label{sec:conclusion}

This work presents the first open-source astronomy package to do accurate astrometry of non-sidereally tracked NEO images with significant elongation in reference stars --- \sw{Astreaks}. We use a novel method to perform astrometry on the non-sidereally tracked images. The pipeline uses a background estimation technique that appropriately captures the background variations while also considering the presence of elongated sources in the field. Further, we methodically compute the SSF of elongated sources by leveraging the knowledge of the telescope system PSF and observation parameters of the target. The pipeline has been designed to meet the goal of accurate astrometry with the novel technique of image segmentation and source de-blending on the background-subtracted image. The catalog of sources detected in this image is used to generate a synthetic image that imitates a sidereal image, had it been taken at the mid-time of the exposure. The astrometry solution of this synthetic image gives the astrometric measurements of the minor planet.

We validate the performance and results of the astrometric accuracy of \sw{Astreaks} on non-sidereally acquired data with GIT using two different instrument set-ups. We achieved an astrometric accuracy of 0.52\arcsecs for 115 NEOs from 371 images, which had stellar streaks up to 6.5$^\prime$. This is well below the 2\arcsecs threshold considered acceptable by the Minor Planet Centre.
% \footnote{Astrometry requirements: \url{https://minorplanetcenter.net/iau/info/Astrometry.html}}
The astrometric results have been tested against the widely used software, \sw{Astrometrica}, where we observe that \sw{Astreaks} outperforms it in detecting elongated reference stars as well as in the accuracy of the astrometric solution for non-siderally tracked images. A comparison of astrometry accuracy revealed that the performance of \sw{Astreaks} on non-sidereal data is as good as point source astrometry in sidereal data.
The modular nature and validation of the pipeline on two different set-ups emphasize that the pipeline can be integrated with other set-ups with simple tweaks in hyperparameters.
%The \sw{Astreaks} python package is publicly available at \url{https://github.com/krittisharma/astreaks} for reproducibility and to foster further research in this domain.

% VB: removed the figure and discussion. Awaiting development of time machine by Kritti to remove figure from GIT paper instead.

% \begin{figure}
% \centering
% \includegraphics[width = \columnwidth]{distribution_of_neos_observed_so_far.png}
% \caption{An ($a$, $e$, $i$) orbital distribution of all solar system objects observed with GIT. The region between perihelion $q = 1.3$~AU and aphelion $Q = 0.983$~AU demarcates the NEOs regime. The periodic comets with $a > 4$~AU and non-periodic comets are excluded from this figure.}
% \label{fig:orbital_distribution}
% \end{figure}

% VB: removed the figure and discussion. Awaiting development of time machine by Kritti to remove figure from GIT paper instead.
%\sw{Astreaks} is being used regularly on GIT for scientific objectives like NEOs and comets follow-up, GROWTH coordinated observations and confirmation of cometary outbursts to produce prolific scientific results. Figure~\ref{fig:orbital_distribution} shows the ($a,~e,~i$) orbital distribution of all minor planets observed with GIT, comprising 18 main belt asteroids, 26 comets and 115 NEOs. The main belt asteroids were primarily observed with the motive of MPC Observatory Code acquisition. We have followed up on 115 NEOs and 26 comets to lower their orbital uncertainties~\footnote{For a complete list of our published astrometric measurements, please refer to our website: \url{https://sites.google.com/view/growthindia/results/asteroids/}}.

\sw{Astreaks} is being used regularly on GIT for the analysis of asteroid data. Some results have been presented in \citet{2021EPSC...15..378S}. In the future, we aim to further improve the astrometric accuracy of \sw{Astreaks} by accounting for the acceleration of NEOs when estimating the streak length for SSF model generation. Typically, the apparent rate of motion of NEOs is between $10^{-1}$ -- $10^{2}$~arcsec/s with accelerations varying between $10^{-7}$ -- $10^{-2}$~arcsec/s$^2$ ~\citep{2012PASP..124.1197V}. 
For a NEO moving with an apparent angular velocity of $10^{2}$~arcsec/s and an acceleration of $10^{-2}$~arcsec/s$^2$, the change in streak length during a 100~s exposure will be by 1\arcsecs. This is a significant change to impact our astrometric accuracy for such long-streaked objects, thus pointing towards the scope of further improvement in the astrometric measurement by \sw{Astreaks}. In addition to astrometry, there is significant scope of improvement in photometric accuracy of \sw{Astreaks} (work in progress). Secondly, we wish to eliminate even the small manual step of identifying the approximate NEO location in the image, which is needed for centroiding and PSF fitting for astrometry and photometry. Lastly, the image segmentation process involves a convolution operation, which is an O(N$^2$) operation for an N$\times$N image, due to which the current implementation of Astreaks is time-consuming ($\sim$ minutes on a generic desktop). Therefore, for faster data processing and submission of observations, we will attempt to make the aforementioned step more time efficient in subsequent versions of the pipeline.

\section*{Acknowledgments}

The GROWTH India Telescope (GIT) is a 70-cm telescope with a 0.7-degree field of view, set up by the Indian Institute of Astrophysics and the Indian Institute of Technology Bombay with support from the Indo-US Science and Technology Forum (IUSSTF) and the Science and Engineering Research Board (SERB) of the Department of Science and Technology (DST), Government of India. It is located at the Indian Astronomical Observatory (Hanle), operated by the Indian Institute of Astrophysics (IIA). We acknowledge funding by the IITB alumni batch of 1994, which partially supports operations of the telescope. Telescope technical details are available at \url{https://sites.google.com/view/growthindia/}.

Kritti Sharma thanks Michael S. P. Kelley (U. Maryland) for his valuable guidance in comet photometry. Kritti Sharma thanks Kunal Deshmukh (IITB) for his guidance during sidereal observations.
Harsh Kumar thanks the LSSTC Data Science Fellowship Program, which is funded by LSSTC, NSF Cybertraining Grant \#1829740, Brinson Foundation, and Moore Foundation; his participation in the program has benefited this work.
B.T.B. is supported by an appointment to the NASA Postdoctoral Program at the NASA Goddard Space Flight Center, administered by Oak Ridge Associated Universities under contract with NASA.
This research has made use of data and/or services provided by the International Astronomical Union's Minor Planet Center (\url{https://www.minorplanetcenter.net/iau/mpc.html}). This work has made use of the \sw{find\_orb}~\citep{findorb} software supplied by Project Pluto (\url{https://www.projectpluto.com/find\_orb.html}). This research has made use of \sw{Astrometrica}~\citep{astrometrica} software (\url{http://www.astrometrica.at/}). This research has made use of the \sw{VizieR} catalogue access tool, CDS, Strasbourg, France.
(DOI:10.26093/cds/vizier). The original description of the VizieR service was published in 2000, A\&AS 143, 23. This research has made use of \sw{Astropy}~\citep{2013A&A...558A..33A, 2018AJ....156..123A}, \sw{NumPy}~\citep{2020Natur.585..357H}, \sw{SciPy}~\citep{2020NatMe..17..261V}, \sw{Matplotlib}~\citep{2007CSE.....9...90H}, \sw{Astro-SCRAPPY}~\citep{2019ascl.soft07032M}, \sw{SExtractor}~\citep{bertin11}, \sw{PSFEx}~\citep{psfex} and \sw{astrometry.net}~\citep{Lang_2010} software. This research has made use of NASA's Astrophysics Data System.

%%%%%%%%%%%%%%%%%%%%%%%%%%%%%%%%%%%%%%%%%%%%%%%%%%%%%%%%%%%%%%%%%%%%

\section*{Data Availability}
\sw{Astreaks} has been made available to the community under an open source license.

\bibliographystyle{mnras}
\bibliography{ref} 

%%%%%%%%%%%%%%%%%%%%%%%%%%%%%%%%%%%%%%%%%%%%%%%%%%%%%%%%%%%%%%%%%%%%

\label{lastpage}
\end{document}